\documentstyle[epsfig]{aa}
\begin{document}

\newcommand{\gsim}{\hbox{\rlap{$^>$}$_\sim$}}
%
   \title{Jetted GRBs, Afterglows and SGRs from Quark Stars Birth}

   \author{Arnon Dar}

   \institute{Theory Division, CERN, CH-1211 Geneva 23 and
Department of Physics, Technion,  
               Haifa 32000, Israel}  
   \date{Received December 20, 1998; accepted} 
   \maketitle
\begin{abstract}
Recent studies suggest that when cold nuclear matter is compressed to high
nuclear densities, diquarks with spin zero and antisymmetric color wave
function Bose condensate into a superfluid/superconducting state that is
several times as dense. Various astrophysical phenomena may be explained
by gravitational collapse of neutron stars (NSs) to (di)quark stars (QSs) 
as a result of a first order phase transition in NSs  within
$\sim 10^{4}$ years after their birth in supernova explosions, when they
cooled and spun down sufficiently (by magnetic braking ?).  The
gravitational energy release drives an explosion which may eject both
highly relativistic narrowly collimated jets and a mildly relativistic
``spherical'' shell. The slow contraction/cooling of the remnant QSs can
power soft gamma ray repeaters (SGRs) and anomalous X-ray pulsars (AXPs),
without invoking a huge magnetic energy storage. The jets can produce the
observed gamma ray bursts (GRBs) in distant galaxies when they happen to
point in our direction and the GRBs afterglows. Also the expanding shell
from the explosion may produce a GRB afterglow. The jets distort the
original SNR, sweep up ambient matter along their trajectories, accelerate
it to cosmic ray (CR) energies and disperse it in hot spots which they
form when they stop in the galactic halo. Such events in our Galaxy may be
the main source of Galactic cosmic rays at all energies. 

\end{abstract}
\section{First Order Phase Transition in Cooling NSs ?} The properties of
hadronic matter under extreme conditions is subject to intense theoretical
studies and experimental investigations in high energy heavy ion
collisions. Numerical techniques cannot yet simulate the behavior of
quarks at very high densities, but, theoretical studies (Alford et al.
1998; Rapp et al. 1998; Wilczek 1998) suggest that in cold $({\rm
T<T_c\sim 50~MeV})$ and dense nuclear matter $({\rm n> 1~fm^{-3}})$ up and
down quarks produce Cooper pairs with spin zero and antisymmetric color
wave function which Bose condensate into a superconducting/superfluid
ground state.  The heat release in energetic heavy ion collisions limits
the experimental information on dense nuclear matter to very high
temperatures.  Information on the behavior of cold nuclear matter at very
high densities must be extracted from observations of neutron stars. Here,
I outline a possible connection between a first order phase transition of
nuclear matter to diquark matter in NSs, GRBs, SGRs and galactic CRs.
\section{Soft Gamma Ray Repeaters}
SGRs are slowly rotating (${\rm P\sim 5-8~s}$)  neutron stars that produce
multiple bursts of soft $\gamma$-rays, often at super-Eddington
luminosities (see, e.g., Hurley and Kouveliotou in these proceedings). 
Four of these objects have been discovered. Two of these objects (SGR
0526-66 and SGR 1900+14) have also produced hard and extremely intense
bursts ($10^{44.5}$ erg in a few tenths of a second, assuming isotropic
emission). The period, P=5.16 s, and the period derivative, ${\rm \dot
P=1.1\times 10^{-10}}$, of SGR 1900+14 (see, e.g., Hurley et al. 1998c and
references therein) yield a characteristic age, ${\rm \tau= P/2\dot
P\approx 750~y}$, much younger than the age of its nearby supernova
remnant (SNR) G42.8+0.6. If this characteristic age of 1900+14 represents
its true age and if it was born at the center of the SNR it must have
moved with perpendicular velocity ${\rm v_\perp\sim
40000(D/5.7~kpc)~km~s^{-1}}$ to its present location (Hurley et al.
1998a; 1998b). The period, P=7.47 s, and the period derivative, ${\rm \dot
P=8\times 10^{-10}}$, of SGR 1806-20 (Kouveliotou et al. 1998), yield a
characteristic age, ${\rm \tau= P/2\dot P\approx 1500~y}$, much younger
than the age of the nearby supernova remnant (SNR) G10.0-0.3. If this age
of 1806+20 represents its true age and if it was born at the center of the
SNR, it must have moved with a perpendicular velocity ${\rm v_\perp\approx
5500(D/14.5~kpc)~km~s^{-1}}$, to its present location (Kulkarni 
1994). If the characteristic age of SGR 0526-66 in SNR N49 in the LMC,
which has a period of ${\rm P\sim 8~s}$, is similar to that of SGR
1900+14, i.e., ${\rm \tau\sim 750 ~y}$, than it is also much younger than
the age of SNR N49, and if it was born at the center of the N49 in the
LMC, it must have moved with a perpendicular velocity ${\rm v_\perp\sim
5000-20000~ km~s^{-1}}$ to its present location (Cline et al. 1982). These
velocities are larger than those of young pulsars, such as the Crab and
Vela pulsars, and those of millisecond pulsars (e.g., Toscano et al 1998)
by about two orders of magnitude.  A similar situation exists for
anomalous X-ray pulsars (AXPs). 

The inferred ages and velocities of SGRs and AXPs suggest that their
characteristic spin down ages are not their true ages. I propose 
that they are the characteristic times that these pulsars spent in
their present phase, a young (di)quark star (QS) that was formed recently,
together with a GRB, by a first order phase transition in neutron stars
(NSs) that have cooled and spun down sufficiently.  Their typical natal
kick velocities are ${\rm v\sim 500~km~s^{-1}}$, much larger than the
typical velocities, $({\rm \sim 100~km~s^{-1}})$, of young neutron stars
produced in SNe, such as the Crab (Caraveo and Mignani 1999) and the Vela 
(Nasuti et al. 1997) pulsars and of all
millisecond pulsars (Toscano et al.  1998). Because of their high natal
velocities, they are expected to be found alone or in wide binaries, but
not in close binaries. Their slow rotation may result from concentration
of their angular momentum in internal superfluid vortices which are pinned
to their crust and brake its rotation. The quiescent thermal X-ray
emission from the cooling QS may be used to verify that their radii are
significantly smaller than those of millisecond pulsars. Finally, the QS
end up as slowly rotating normal radio pulsars.  
\section{Highly Relativistic Jets From NS Collapse}
Relativistic jets seem to be emitted when mass is accreted at a high rate
onto compact objects. They are observed in galactic superluminal sources,
such as the microquasar GRS 1915+105 (Rodriguez and Mirabel 1994), where
mass is accreted onto a stellar black hole (BH), in star binaries
containing 
NSs such as SS433 (e.g., Hjellming \& Johnston 1988), and in many active
galactic nuclei (AGN), where mass is accreted onto a supermassive BH. The
ejection of the relativistic jets is not well understood yet.  But, it
seems very likely that highly relativistic jets are also ejected
in NSs collapse because the accretion rates and the magnetic fields
involved in such collapse are much larger. We estimate their kinetic
energy ${\rm E_{jet}}$ and bulk motion Lorentz factor $\Gamma$ as follows: 

Most of the gravitational energy release in the gravitational collapse of
NS (${\rm M\sim~1.4M_\odot } )$ to QS, is radiated in a short burst of
neutrinos and only a small part of it is released as kinetic energy of the
remnant QS and of the ejecta which probably consists of a mildly
relativistic spherical shell and highly relativistic two antiparralel jets. 
If momentum imbalance in the ejection of the relativistic jets ( and not
asymmetric neutrino emission)  is responsible for the observed large mean
velocity (Lyne and Lorimer 1994)  $v\approx 450\pm 90~ {\rm km~s^{-1}}$,
of slowly spinning pulsars (presumably DQ), then momentum conservation
implies that the kinetic energy of the jets satisfies \begin{equation}
{\rm E_{jet} \geq cP_{ns}\sim vM_{ns}c\sim 4\times 10^{51}~erg}. 
\end{equation} If two antiparallel jets are ejected, then the jet kinetic
energy may be of the order of ${\rm E_{jet}\sim 10^{52}~erg}$. 

High-resolution radio observations have resolved narrowly collimated
relativistic jets from microquasars, quasars and radio galaxies into
clouds of plasma (plasmoids)  that are emitted in injection episodes which
are correlated with sudden removal of the accretion disk material. In GRS
1915+105 these plasmoids appear to expand freely in the plasmoid rest
frame, (i.e., with the speed of sound in a relativistic gas, ${\rm c_s\sim
c/\sqrt{3}}$, corresponding to vertical expansion speed of ${\rm \sim
c/\Gamma\sqrt{3}}$ in the lab frame), up to a radius (Rodriguez and Mirabel 
1998) ${\rm R_ p\sim 2\times 10^{15}~cm}$ and retain a constant or slowly
expanding radius afterwards, probably, due to magnetic confinement.
Likewise plasmoids from quasars seem to retain a constant radius after
initial expansion. The highly relativistic magnetized plasmoids from NSs
collapse are strong emitters of beamed $\gamma$-rays through synchrotron
emission, inverse Compton scattering and resonance scattering of
interstellar light.  When they point in our direction from external
galaxies, they can produce the observed GRBs and their afterglows (see
next section).  If the true rate of GRBs is similar to the birth rate of
NSs, ${\rm \dot N_{ns}\simeq 2\times 10^{-2}~ y^{-1}}$ in galaxies like
our own (e.g., van den Bergh and Tamman 1991), then, since the estimated
current
rate of GRBs in galaxies like our own is (e.g., Wijers et al. 1997)  ${\rm
\dot N_{grb}\sim 10^{-8}~y^{-1}}$, it implies that the radiation emitted
in GRBs must be narrowly beamed into a solid angle $\Delta\Omega$ that
satisfies
\begin{equation} 
{\rm \Delta\Omega\simeq 2\pi \dot N_{grb}/\dot
N_{ns}\simeq \pi\times 10^{-6}}. 
\end{equation} 
GRB observations and the recent redshift measurements/estimates of some
GRBs and of host galaxies of GRBs (see this volume) do suggest that GRBs
are produced by highly relativistic, narrowly collimated jets (Shaviv and
Dar 1995; Dar 1998 and references therein) with bulk motion velocities $v$
corresponding to Lorentz factors $\Gamma\sim 10^3$ (Baring and Harding
1997). Emission from narrow jets with
$\Gamma=10^3$, indeed, is beamed into $\Delta\Omega\sim
\pi/\Gamma^2\simeq\pi\times 10^{-6} $ consistent with eq. 2.  Such strong
beaming implies that we observe only a very small fraction of a large rate
of the events that can produce GRBs. We call these events ``Galactic''
GRBs (GGRBs)  if they occur in our Milky Way (MW) galaxy and Cosmological
GRBs (CGRBs) if they occur in distant galaxies.  Eqs. 1,2 also imply that
the ejected jet (plasmoid) has a mass ${\rm M_{jet}\sim 1.5\times
10^{-6}M_{ns}\sim 2.1\times 10^{-6}M_\odot\sim 0.7M_{Earth}}$. Even if
only a fraction $\eta\sim 10^{-5}$ of the jet kinetic energy is radiated
in $\gamma$-rays, the inferred ``isotropic'' $\gamma$-ray emission in GRBs
is ${\rm E_{isot}\simeq 4\eta \Gamma^2 E_{jet}\sim 4\times 10^{53}~erg}$,
while the true $\gamma$-ray emission is only ${\rm E_\gamma\sim
10^{47}~erg}$.
\section{Jetted GRBs} 
Observations of GRB afterglows suggest that GRBs are produced in star
burst regions (SBR)  where the column  densities of gas and light
are high. The ejected jets consist of pure ${\rm e^+e^-}$ plasmoids or of
normal hadronic gas/plasma clouds. The GRB can be produced by electron
synchrotron emission (for various GRB scenarios based on synchrotron
emission see papers in this volume and references therein). If the jet
consist of a single plasmoid, then individual $\gamma$-ray pulses that
make GRBs may be produced by magnetic shocks induced by internal
instabilities or by external inhomegeneities.  If the jet consists of
multiple ejections of plasmoids, then the GRB pulses may be produced when
later ejected plasmoids collide with earlier ejected plasmoids that have
been slowed down by sweeping up the interstellar medium in front of them.
Perhaps, Early time variability of GRBs afterglows may be used to
distinguish between the single and multiple ejection scenarios. However,
these scenarios do not seem to provide a simple explanation for the
$\sim$ MeV ``peak energy'' of GRBs. Other emission mechanisms do provide
such an explanation:

If the highly relativistic plasmoid consists of a pure ${\rm e^+e^-}$
plasma, then inverse Compton scattering of stellar light (${\rm 
h\nu=\epsilon_{eV}\times 1~eV}$) by the plasmoid can explain the observed
typical $\gamma$ energy (${\rm \epsilon_{\gamma}\sim 4\Gamma_3^2
\epsilon_{eV} /3(1+z)~MeV}$), GRB duration ($ {\rm T\sim
R_{SBR}/ 2c\Gamma^2\sim 50~~s}$), pulse duration (${\rm t_p\sim R_p/
2c\Gamma^2\sim 150~ms}$), fluence $({\rm F_\gamma\sim
10^{-5}}$${\rm ~erg~cm^{-2}})$, light curve and spectral evolution 
of GRBs (Shaviv and Dar 1995; Shaviv 1996; Dar 1998). For instance,
\begin{equation}
{\rm  F_\gamma \simeq{\sigma_{_T} N\epsilon_\gamma\over \Gamma m_ec^2}
{E_{jet}(1+z)\over D^2\Delta\Omega}\simeq { 10^{-5}z_2 N_{22}\Gamma_3
\epsilon'_\gamma E_{52} \over D_{29}^2}~{erg\over cm^2}}  
\end{equation}
where ${\rm D=D_{29}\times 10^{29}~cm}$ is the luminosity distance of the
GRB at redshift z, ${\rm z_2=(1+z)/2}$, ${\rm N=N_{22}\times
10^{22}~cm^{-2}}$ is the column density of photons along the jet
trajectory in the star burst region, ${\rm \sigma_{_T}=0.65\times
10^{-24}~cm^2}$ is the Thomson cross section, ${\rm E_{jet}=E_{52}\times
10^{52}~erg}$ and $\Gamma=\Gamma_3\times 10^3$. 

If the plasmoid consists of normal crust material of neutron stars,
(Doppler shifted ${\rm K_\alpha}$ iron line was detected from the jets of
SS433), than photoabsorption of stellar light by partially ionized iron
(Doppler shifted to X-rays in the jet rest frame) and its reemission as
$\gamma$ rays (iron X-rays lines in the jet rest frame) yield ${\rm
\epsilon_\gamma=\Gamma\epsilon_x/(1+z)\sim MeV}$) in the observer frame 
and
\begin{equation}
{\rm  F_\gamma \simeq{\sigma_{a} N\epsilon_\gamma\over \Gamma M_{Fe}c^2}
{E_{jet}(1+z)\over D^2\Delta\Omega}\simeq
{10^{-5} z_2\sigma_{19} N_{22}\epsilon_x\Gamma_3E_{52} \over
D_{29}^2}~{erg\over ~cm^2}}
\end{equation}
where ${\rm \sigma_a= \sigma_{19}\times 10^{-19}~cm^2}$ is the mean
photoabsorption cross section of X-rays by partially ionized iron.  
\section{GRB Afterglows}
The afterglows of GRBs may be synchrotron emission from the decelerating
plasmoids (e.g., Chiang and Dermer 1997), and then they are highly beamed
and may exhibit detectable superluminal velocities (${\rm v_\perp\leq
\Gamma c}$ during and right after the GRB).  The deceleration of a mildly
relativistic spherical ejecta from the NS collapse to QS, may also produce
additional (stationary) afterglow (many planetary nebulae, e.g., NGC 7009,
NGC 6826, and some SNR look like ejection of antiparralel jets from a
spherical explosion).  When transformed to the rest frame of the
decelerating plasmoid, the synchrotron spectra can be modeled by
convolving typical electron energy spectrum (${\rm E^{-p}}$ at low
energies up to some ``break energy'' where it steepens to ${\rm E^{-p-1}}$
and cutoffs exponentially at some higher energy due to synchrotron losses
in Magnetic acceleration)  with the synchrotron Green's function (see,
e.g., Meisenheimer et al. 1989). In the observer frame it yields spectral
intensity (Dar 1998) 
\begin{equation}
{\rm I_\nu\sim \nu^{\alpha}t^{\beta}\sim\nu^{-0.75\pm 0.25} t^{-1.25\pm 0.08}},
\end{equation}
where ${\rm \alpha =-(p-1)/2}$ and ${\rm \beta=-(p+5)/6}$ and where I
assumed  ${\rm p=2.5\pm 0.5}$ for Magnetic Fermi acceleration. This
prediction
is in agreement with observations of GRB afterglows. Moreover, the glows
of microquasar plasmoids and radio quasar jets after ejection and of
blazar jets after flares show the same universal behaviour as observed in
GRBs afterglows. For instance, the glows of the ejected plasmoids from
GRS 1915+105 on April 16, 1994 near the source had $\alpha=-0.8\pm 0.1$ and
$\beta= - 1.3\pm 0.2$ (Rodriguez and Mirabel 1998) identical to those
observed for SS 433 (Hjellming \ Johnston 1988) and for the inner regions
of jets of some radio galaxies (Bridle \& Perley 1984). 
\section {GGRBs - The Source of Cosmic Rays}  
According to  the current paradigm  of CR origin, CR nuclei with energies
below 3 $\cdot$ 10$^{15}$ eV (the ``knee'') are accelerated in Galactic
SNRs (Ginzburg 1957) and those above $3\times 10^{18}$
eV (the ``ankle''), for which a disk origin is unlikely due to their
isotropy, in sources far beyond our Galaxy (Burbidge 1964). However,
recent observations, in
particular the absence (Takeda 1998) of the ``GZK cutoff'' in the CR
intensity at energies above $\sim 10^{20}$ eV due to interactions with the
cosmic microwave radiation (Greisen 1996; 
Zatsepin \&  Kuz'min 1966) have put into question (e.g., Hillas 1998)
the current paradigm for the CR origin.  Dar and Plaga
(1998) have recently proposed an alternative paradigm where
Galactic GRBs (GGRBs) are the main source of the CRs at all energies
and no GZK cutoff is expected:  

Relativistic jets are efficient CR accelerators (e.g., Dar 1998b).  The
highly relativistic narrowly collimated jets/plasmoids from the birth or 
collapse of NSs in the disk of our Galaxy
that are emitted with ${\rm E_{jet}\sim 10^{52}~ erg}$ perpendicular to
the Galactic disk, stop only in the Galactic halo, when the rest mass
energy of the swept up ambient material becomes $\geq$ their initial
kinetic
energy. Through  the Fermi mechanism, they accelerate the swept up ambient
matter to CR energies and disperse it into the halo from the hot spots
which they form when they finally stop in the Galactic halo (Fig.1).
The typical equipartition magnetic fields in such hot spots may reach
${\rm B\sim (3E_{jet}/R_ p^3)^{1/2}\sim 1~ G}$. Synchrotron losses cut off
Fermi acceleration of CR nuclei with mass number A at ${\rm E\sim \Gamma
A^2Z^{-3/2}(B/G)^{-1/2}\times 10^{20}~eV.}$ Particle-escape cuts off Fermi
acceleration when the Larmor radius of the accelerated particles in the
plasmoid rest frame becomes comparable to the radius of the plasmoid,
i.e., above $ {\rm E\simeq \Gamma Z(B/G)(R_p/0.1~pc)\times 10^{20}~eV}.$
Consequently, CR with ${\rm E>Z\times
10^{20}~eV}$ can no longer be isotropized by acceleration or deflection in
hot spots with $\Gamma\sim 1$. Thus,  according to Dar and Plaga (1998)
CR with energies above ${\rm 10^{20}~eV}$, probably, are heavy nuclei.

Fermi acceleration in or by the highly relativistic jets from GRBs can
produce a broken power-law spectrum, ${\rm dn/dE\sim E^{-\beta}}$, with
$\beta\sim 2.2$ below a knee around ${\rm E_{knee}\sim A~ PeV}$ and
$\beta\sim 2.5$ above this energy (Dar 1998b). Spectral indices $\beta$
$\sim 2.2$ were obtained also in numerical simulation of relativistic
shock acceleration (e.g., Bednarz and Ostrowski 1998) . Galactic
magnetic confinement increases the density of Galactic CR by the ratio
${\rm c\tau_{\rm h}/R_G}$ where ${\rm \tau_h(E)}$, is the mean residence
time in the halo of Galactic CR with energy E and ${\rm R_G\sim 50~ kpc}$
is the radius of the Galactic magnetic-confinement region. With the
standard choice for the energy dependence of the diffusion constant
(observed e.g., in solar-system plasmas) one gets:
${\rm\tau_h\propto (E/Z)^{-0.5}}$.  Consequently, the energy spectrum of
CR is predicted to be
\begin{equation} {\rm 
dn/dE\sim C (E/E_{ knee})^{-\alpha}} 
\end{equation} 
with $\alpha\simeq \beta+0.5\simeq 2.7~ (\simeq 3)$ below (above) the
knee.  This power-law continues as long as the Galactic magnetic field
confines the CRs.

Part of the kinetic energy released by GGRBs is transported into the Galactic
halo by the jets. Assuming equipartition of this energy, without large
losses, between CR, gas and magnetic fields in the halo during the
residence time of CR there, the magnetic field strength  
B$_{\rm h}$ in the halo is
expected to be comparable to that of the disk ${\rm B_ h \sim
(2L_{\rm MW}[CR]\tau_ h /R_ h^3)^{1/2}\simeq 3~\mu G}$
where ${\rm \tau_ h\sim 5\times
10^9~ y}$ is the mean residence time of the bulk of the CR in the Galactic
halo. Cosmic rays with Larmor radius larger than 
the coherence length $\lambda$  of the halo magnetic fields, i.e., with
energy above  
\begin{equation} {\rm
E_{ankle}\sim 3\times 10^{18}(ZB_h/3\mu G)(\lambda/kpc)~eV}, 
\end{equation}
escape Galactic trapping. Thus, the
CR ankle is explained as the energy where the mean residence time
${\rm \tau_h(E)}$ of CR becomes comparable to the free escape time from
the halo
$\tau_{\rm free}\sim 1.6(R_{\rm h}/50~{\rm kpc})\times 10^5~{\rm years}$.
Therefore, the spectrum of CR with energies above the ankle, that do not suffer Galactic
magnetic trapping, is the CR spectrum produced by the jet, i.e.,
\begin{equation} 
{\rm
dn/dE\sim C (E_{ankle}/E_{knee})^{-3}(E/E_{ankle})^{-2.5};~
E> E_{ankle}}.
\end{equation}
Eqs. 6,7,8 describe well the overwhole CR energy spectrum.

\begin{figure}[thb] 
\epsfig{file=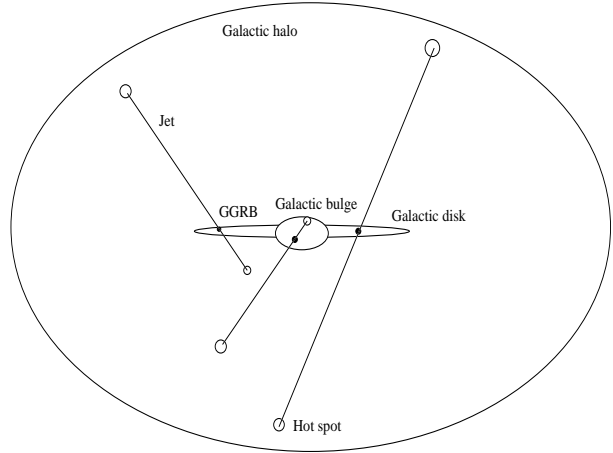,width=8cm,height=6cm,clip=,angle=0}
\caption{\label{fig1} \it A highly schematic sketch of the Dar-Plaga 
paradigm. The birth of QS or the
collapse of NSs in the disk of our Galaxy leads to an
ejection of two opposite jets that produce ``hot spots'' when they stop in
an extended Galactic halo.}
\end{figure} 
\begin{acknowledgements} Results presented here are based on
an on going collaboration with R. Plaga. Useful comments by N. Antoniou
and A. DeR\`ujula are gratefully acknowledged. 
\end{acknowledgements}

{}

\end{document}